\documentclass[showpacs,preprintnumbers,amsmath,amssymb,aps, prd]{revtex4}
\usepackage{graphicx}
\usepackage{grffile}
\usepackage{caption}
\usepackage{subcaption}
\usepackage{float}
\usepackage{color}
\usepackage{dcolumn}
\usepackage{bm}
\begin{document}
\title{Thermodynamics of a Schwarzschild-like black hole with a minimum observable length
and the radiation process of a thin accretion disc around it}
\author{Himangshu Barman}
\affiliation{Hooghly Mohsin College, Chinsurah, Hooghly,
PIN-712101, India} \email{himu.phy@gmail.com}
\author{Mohamed Moussa}
\affiliation{Physics Department, Faculty of Science, Benha
University, Benha 13518,
Egypt}\email{mohamed.ibrahim@fsc.bu.edu.eg}
\author{Homa Shababi}
\affiliation{Center for Theoretical Physics, College of Physical
Science and Technology, Sichuan University, Chengdu 610065, P. R.
China}\email{h.shababi@scu.edu.cn}
\author{Anisur Rahaman}
\email{anisur.associates@aicuua.in;
manisurn@gmail.com(Corresponding Author)} \affiliation{Durgapur
Government College, Durgapur-713214, India}
\date{\today}
\begin{abstract}
We study quantum gravity effects on the thermodynamic character
and the  radiation process of the thin accretion disks around
Schwarzschild-like black hole. The quantum gravity correction is
invoked through the framework of generalization of uncertainty
which is equivalent to the renormalization group improved quantum
gravity and maintain the limit of the asymptotically safe
preposition of gravity. It admits a free parameter that encodes
the quantum effects on the spacetime geometry. It allows us to
study how the thermal properties of the black hole itself and the
the accretion around it disk are modified in the quantum regime.
We computed explicitly the entropy, temperature, free energy, and
enthalpy of the modified black hole and show its variation with
with the free parameter that encodes the quantum effects. We
explicitly make estimations of quantum correction to the time
averaged energy flux, the temperature of the disk, the
differential luminosity, and the conversion efficiency of
accreting mass into radiation. We observe a conspicuous shifting
of the radius of the innermost stable circular orbit (ISCO) toward
small values together with  an enhancement of the maximum of the
values of the average thermal radiation and greater conversion
efficiency of accreting mass into radiation compared to the
classical gravity scenario.
\end{abstract}

\maketitle

\section{Introduction}
The black hole is a remarkable solution to the classical field
equations thanks to Einstein within the context of general
relativity theory. In an astrophysical sense, it is the regions of
spacetime that gets deformed undergoing gravitational collapse.
The solutions of the equation are inflected with singularities
inside the event horizon. Therefore it loses its prophetic
character and it is no longer possible to extend classical
formulation of spacetime inside the event horizon. This forces to
think about general relativity as an effective theory of gravity
that remains valid solely up to certain energy scales. At high
energy scales, such as the Planck scale, it is expected that a
full theory of quantum gravity will resolve the unphysical
singularities in the spacetime manifold. So the predictively be
rehabilitated. However, attempts of describing gravity within the
framework of quantum theory face the dander of the perturbative
non-renormalizable character of general relativity. As a
consequence, different approaches have surfaced, namely a loop
quantum gravity spin foams \cite{ASHTE1, ASHTE2, ASHTE3, ROVEL,
APER}, string theory \cite{GREEN, WITTEN1, WITTEN2}, etc.. Another
promising proposal to deal with this downside is that the
Asymptotic Safety scenario which uses the useful techniques of the
functional renormalization group. The existence of a non-Gaussian
fixed point of the gravitational renormalization group flow that
controls the behavior of the idea at trans-Planckian energies is
that the main speculation within this construction. The physical
degrees of freedom interact predominantly with anti-screening
within the neighborhood of the non-Gaussian fixed point which
renders physical quantities safe from unphysical divergences
within the vicinity of the Plank scale which renders physical
quantities safe from unphysical divergences \cite{AMBOJ}. It is
the fact that asymptotic safety defines a harmonious and
prognostic scientific proposition among the frame of quantum field
theory we couldn't ignore its eventuality, still, it stands as a
vaticination since a rigorous evidence for the existence of the
non-Gaussian fixed point continues to be lacking. It is the fact
that AS defines a consistent and prognostic scientific theory for
gravity among the framework of quantum theory we could not ignore
its potential, however, it stands as a prediction since a rigorous
existence proof for the NGFP continues to be lacking. There is,
however, substantial evidence supporting the existence of the
non-trivial renormalization group fixed point at the center of
this construction \cite{LEE, REU}. The importance of black holes
as testing ground for gravity theories within the strong field
regime has motivated numerous studies on the implications of
asymptotic safety gravity for black hole physics, most of them
geared toward determining quantum corrections to the classical
metrics.

The generalization of uncertainty is a fascinating extension that
has its origin in string theory and loop quantum gravity. In the
article \cite{REU1}, the author analyzes quantum gravity
corrections to the accretion onto black holes within the context
of asymptotically safe gravity. The asymptotic safety was
maintained there by invoking the running Newton's constant
proposition. The identical asymptotic safety could be contemplated
within the Generalize Uncertainty Principle (GUP) framework which
we will attempt to explore and study the physical quantities
connected to the accretion process maintaining AS during this
composition. The usage of generalized scientific theory based on
GUP, which renders asymptotic safety, is used here to account for
quantum gravity corrections to the Schwarzschild metrics. Quantum
spacetime with running gravitational coupling that is included
into the scenario gets manifested in effective mass.

Investigating the gravity-induced quantum interference pattern,
followed by the Experiment with Gedanken-experiment to determine
the weight of photon it  has been established  that the running
Newton's law applies to photons. Examining the quantum
interference pattern caused by gravity, and then the thought
Gedanken-experiment  for determining the weight of a photon,  it
has been established that the running Newton gravitational
constant can be stimulated by the principles of generalized
uncertainty, which  ends up in quantum gravity corrections to
Schwarzschild region metrics \cite{BKO}. The enhanced
quantum-corrected metric is used here cherish the Schwarzschild
metric indeed and will be used to study the quantum-corrected
thermodynamics of the black hole and the thermal radiation process
of thin accretion disc around the black hole.

The paper is organized as follows. In Sec. II we amend quantum
gravity correction to the the Schwarzschild metric by the use of
GUP and study the impact of GUP on the nature of horizon of this
GUP inspired quantum gravity corrected Schwarzschild metric. In
Sec. III we describe the the thermodynamical characteristics of
this modified black hole. In  Sec. IV is devoted to the
Description of the geodesic in the GUP inspired quantum corrected
spacetime geometry. In Sec. V, we study the impact of GUP
parameter in Mass accretion rate and differential luminosity for
the thin accretion Sec. VI contains a brief summary and
discussion.
\section{Insertion of quantum gravity effect into the Schwarzschild metric}
Before jumping into the formulation of having a modified black
hole endowed with quantum gravity effect through GUP perspective
it would be beneficial to give a brief description of GUP. Let us
turn into that.
\subsection{Description GUP with minimum measurable length}
In recent times it has been noticed that various approaches to
quantum gravity including string theory \cite{VEN, AMIT, GROSS,
KONISH, KATO, GUIDA}, noncommutative geometry \cite{KATO}, loop
quantum gravity \cite{GREY} predict the existence of a minimum
measurable length of the order of Planck length. It leads to
different generalizations of usual uncertainty relation in the
context of quantum gravity\cite{GREY, FELDER, GROSS, AMIT1,
WITTEN}, although  Heisenberg uncertainty principle is the
cornerstone of the formulation of quantum theory that puts a
fundamental limit on the precision of measuring the position and
momentum. In the  Heisenberg uncertainty principle,  $\Delta x \to
0$ leads to $\Delta p \to \infty$, therefore,  the standard
Heisenberg uncertainty relation $\Delta x \Delta p \ge \hbar$
becomes skimpy to explain the existence of a minimum measurable
length. Thus, it necessitates the replacement of  Heisenberg
uncertainty principle by the Generalized Uncertainty Principle
(GUP) \cite{KEMPF} to accommodate the possibility of minimum
measurable length. In the article \cite{KEMPF} Kempf showed that a
generalized uncertainty relation could be defined by
\begin{eqnarray}\label{GUP}
 \Delta x_{k} \Delta p_{l}\geq \hbar\delta_{kl}\left(1+\eta
\left[\left(\Delta p\right)^{2}+ {\left\langle p\right\rangle
}^2\right]\right),
\end{eqnarray}
where $\eta$ is GUP parameter which has the  the definition
$\eta=\frac{\eta_o}{M_{PL}c^2}$, where $M_{PL}$ is the Planck mass
and $\eta_0$ is parameter of the order of unity.  The expression
(\ref{GUP}) has the potential to  accommodate the minimum
measurable length within the revised principle. It is
straightforward to see that for the above GUP (\ref{GUP}), the
minimum non-zero length is found out to be
\begin{eqnarray}
(\Delta x)_{Min}=\hbar\sqrt\eta\sqrt{1+\eta{\left\langle
p\right\rangle}^2},
\end{eqnarray}
where setting $\left\langle p\right\rangle=0$ results to the
absolute minimal measurable length:
\begin{eqnarray}
(\Delta x)_{Min}=\hbar\sqrt\eta=\sqrt{\eta_0}l_{PL},\label{MIN}
\end{eqnarray}
where  $l_{PL}=(\frac{G\hbar}{c^{3}})^{\frac{1}{2}}\approx
10^{-35}$m \cite{KEMPF, KEMPF1, KEMPF2} is the Planck length.
 This generalized uncertainty (\ref{GUP}) corresponds to the
following deformed commutation relation between position and
momentum \cite{KEMPF}
\begin{eqnarray}
\left [x_{k},p_{l}\right]=i\hbar\delta_{kl}\left(1+\eta
p^{2}\right),\label{GUP1}
\end{eqnarray}
where $p^2=\sum_k p_k^2$. What follows next is an attempt to have
a modification of the relation (\ref{GUP}) and  (\ref{GUP1}) in
one dimension by
\begin{equation}
\Delta x \Delta p \geq \hbar \left(1+\eta (\Delta p)^2\right),
\label{GUN}
\end{equation}
\begin{equation}
[x,p]=i\hbar \left(1+\eta p^2\right)=i\hbar \zeta,\label{G10}
\end{equation}
where $\zeta=1+\eta p^2$. Hence, Eqn. (\ref{GUN}) can be rewritten
as
\begin{equation}
\Delta x \Delta p \geq \hbar \zeta.\label{GUP3}
\end{equation}
\subsection{Schwarzschild metric endowed with quantum gravity
correction}
Let us now formulate a Schwarzschild-like spacetime
metric where quantum gravity correction gets induced through
generalized uncertainty principle keeping in sight on the
Gedanken-experiment initially proposed by Einstein. Einstein made
a trial to exhibit the violation of the indeterminacy principle
through a Gedanken-experiment which was purported to measure the
load of photons \cite{XLING, BOHR, NOV}. He assumed a box
containing photon gas with a totally reflective wall was suspended
by a spring scale. There was a system inside the box that caused
the shutter to open and shut at moment $\tau$ for the time
interval $\Delta t$, which allowed to passing out just one photon.
A clock capable of showing extremely high precision measurement of
your time could be accustomed measure the interval $\Delta \tau$
and at the same time the mass difference of the box would
determine the energy of the emitted photon as per Einstein's
assumption, the amount required for photon radiation is precisely
$\Delta \tau\rightarrow 0$ which may be lead to the violation of
the uncertainty relation for energy and time, i.e. $\Delta E\Delta
\tau\rightarrow 0$. However Bohr argued that \cite{BOHR},
Einstein's deduction wasn't flawless since he neglected the
time-dilation effect which might play a significant role due to
the difference of gravitational potential. Based on general
relativity, when altitude changes, the speed of your time flow
also changes thanks to the change in their gravitational
potential. Thus, for the put down the box, the time uncertainty
$\Delta \tau$, in terms of the vertical position uncertainty
$\Delta x$, would be expressed as \cite{BOHR, NOV}
\begin{equation}
\Delta \tau=\frac{g \Delta x}{c^2}\tau,\label{0}
\end{equation}
where $\tau$ represents the time period of weighing the photon. As
it is known, according to the quantum theorem, the uncertainty
relation in energy and time of the photon is express as $\Delta E
\Delta \tau\geq \hbar$ which after substituting Eqn. (\ref{0})
turns into
\begin{equation}
\Delta E \geq \frac{\hbar c^2}{gt\Delta x}.\label{G1}
\end{equation}
Let us now look at the relation between the weight of the photon
in the Gedanken-experiment and the corresponding quantities in
quantum mechanics in the GUP framework (\ref{G1}). If we focus on
the original position of the pointer on the box before opening the
shutter, we will find that after releasing photon in the box the
pointer moves up with reference to its original position. To get
back the pointer in its original position in a period of time
$\tau$, some weights equal to the weight of the photon must be
added to the box. If we now use Eqn. (\ref{G1}), having accuracy
in measuring the position $\Delta x$ as marked by the indicator of
the clock, the minimum uncertainty in momentum $\Delta P_{Min}$
will be
\begin{equation}\label{2}
\Delta p_{Min}=\frac{\zeta\hbar}{\Delta x}.
\end{equation}
since the quantum weight limit of a photon is $g \Delta m$, so in
a period of time $\tau$ the smallest photon
 weight will be equal to $\zeta\hbar/\tau \Delta x = \Delta p_{Min}  / \tau\leq g\Delta m$.
  Now, using Eqn. (\ref{2}) we also find that
\begin{equation}
\zeta\hbar=\Delta x \Delta p_{Min}\leq g \tau \Delta x \Delta
m.\label{3}
\end{equation}
If  the relation $\Delta E = c^2 \Delta m$, is used for  $\Delta
m$ the  Eqn. (\ref{3}) can be written down as
\begin{equation}
\Delta E\geq \frac{\zeta\hbar c^2}{g\tau\Delta x},\label{4}
\end{equation}
where $\zeta$ refers to the the GUP effects. Note that in the
absence of GUP framework, i.e. when $\eta\rightarrow 0$, the
standard energy-time uncertainty relation Eqn. (\ref{G1}) is
reobtained.

A careful look on the standard uncertainty relation between energy
and time (\ref{G1}) and the generalized uncertainty relation
(\ref{GUP3}) gives rise to an interpretation that gravitational
field strength $g$ is modified to $\frac{g}{\zeta}$. Then,
replacing $g$ by $\frac{g}{\zeta}$ we have
\begin{equation}\label{5}
\bar{g}=\frac{g}{\zeta}=\frac{G_0M}{\zeta R^2}.
\end{equation}
Hence, using (\ref{5}), the modified Schwarzschild metric turns
into
\begin{equation}\label{6}
ds^2=-\left(1-\frac{2G_0M}{\zeta c^2r}\right)c^2dt^2
+\left(1-\frac{2G_0M}{\zeta c^2r}\right)^{-1}dr^2+r^2(d \theta^2 +
sin^2\theta d\phi^2),
\end{equation}
where $G_0$ stands for universal gravitational constant. On the
other hand, as stated in some other literature \cite{XLING,
NARLI}, when two virtual particles with energies $\Delta E$ are at
a distance $\Delta S$ from each other, the tidal force between
them is obtained by
\begin{equation}\label{f1}
F=\frac{2G_0M}{r^3}\frac{\Delta E}{c^2}\Delta x.
\end{equation}
So, the uncertainty in momentum is given by
\begin{equation}\label{f2}
\Delta p=F\Delta \tau=\frac{2G_0M}{r^3}\frac{\Delta E}{c^2}\Delta
x\Delta \tau,
\end{equation}
where $\Delta t$ represents the life time of the particle.

If virtual particles turns into  real particles having the
exposure of tidal force, the uncertainty relations $\Delta p\Delta
x\geq \hbar$ and $\Delta E\Delta t\geq \hbar$ can be used with
reasonably well justifiable manner. Therefore, using these
uncertainty relations in (\ref{f2}), we find that
\begin{equation}\label{f3}
(\Delta p)^2\geq \frac{2\hbar^2G_0M}{c^2r^3}.
\end{equation}
Accordingly, it can be written that $p^2\approx(\Delta p)^2
\approx \frac{2\hbar^2G_0M}{c^2r^3}$. Hence applying this modified
uncertainty relation, we obtain the Schwarzschild metric in the
presence of minimal measurable length as
\begin{equation}
ds^2=-\left(1-\frac{2G_0Mr^2}{c^2\left(r^3+\eta
\frac{2\hbar^2G_0M}{c^2}\right)}\right)c^2dt^2+\left(1-\frac{2G_0Mr^2}{c^2\left(r^3+\eta
\frac{2\hbar^2G_0M}{c^2}\right)}\right)^{-1}dr^2+r^2(d \theta^2 +
sin^2\theta d\phi^2).\label{SCH}
\end{equation}
This metric specifies a set of spacetime that depends on different
scales of momentum via the modified mass of the black hole.  For
dimensionless case, the modified Schwarzschild metric (\ref{SCH})
reads
\begin{equation}
ds^2=-\left(1-\frac{2Mr^2}{r^3+2\eta
M}\right)dt^2+\left(1-\frac{2Mr^2}{r^3+2\eta
M}\right)^{-1}dr^2+r^2(d \theta^2 + sin^2\theta
d\phi^2).\label{SCH1}
\end{equation}
In the following sections, for the sake of simplicity, we use the
modified Schwarzschild metric (\ref{SCH}) in the form of
\begin{equation}
ds^2= -f(r \eta) c^2 dt^2 + \frac{1}{f(r \eta)} dr^2 + r^2(d
\theta^2 + sin^2\theta d\phi^2), \label{SCH2}
\end{equation}
where
\begin{equation}
f(r \eta)=1-\frac{2G_0Mr^2}{c^2\left(r^3+\eta
\frac{2\hbar^2G_0M}{c^2}\right)} =
1-\frac{2MG_0}{c^2r}\frac{r^3}{r^3+2\eta
\hbar^2(\frac{MG_0}{c^2})}=1-\frac{2M}{c^2r}G(r \eta ),\label{F}
\end{equation}
where
\begin{equation}
G(r \eta)= \frac{G_0 r^3}{r^3+2\eta \hbar^2(\frac{MG_0}{c^2})}.
\label{G}
\end{equation}
So for $\eta\to 0$, $G(r \eta) \to G_0$ and the quantum correction
disappears and we get back to the Schwarzschild metric. If we
consider the metric in equation in natural unit setting $c=1$ and
$G_0=1$ in (\ref{SCH1}) to find out the position of the Horizon we
need to have the solution of the equation
\begin{equation}
1-\frac{2Mr^2}{r^3+2\eta M}=0, \label{HOR}
\end{equation}
which has the solution
\begin{equation}
r_H = \frac{2}{3}M +\frac{4}{3}M
cos[\frac{1}{3}cos^{-1}(1-\frac{27}{8}\frac{\eta}{M^2}),
\label{RH}
\end{equation}
provided the mass of the black hole satisfy the condition $M >
M_c$, where $M_c$ is called the critical mass:
$M_c=\frac{27}{16}\eta$. When $M \le M_c$ it fails to describe any
horizon in the spacetime geometry since equation (\ref{HOR}) can
not provide any positive solution in that situation. Plots of the
improved metric coefficient $f(r)$ for different value of mass $M$
with $\eta=\frac{16}{27}$ and $\eta=0$ (classical) are given
below.
\begin{figure}[H]
 \begin{center}
\includegraphics[scale=0.5]{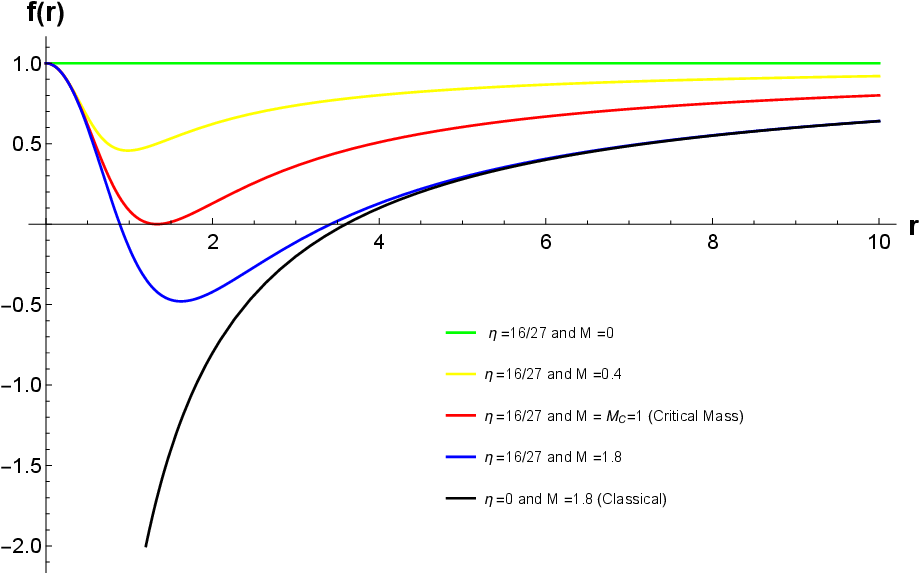}
 \caption{Plots of
the improved metric coefficient $f (r \eta)$ for $M = 0.4 <M_c$
(yellow), $M = M_c$ (red), and $M = 1.8 > M_c$ (blue), with $M_c =
1$ corresponding to $\eta= \frac{16}{27}$. The black line shows
the classical $f_{0}(r)$ for $M = 1.8$ corresponding to $\eta=0$
and green line shows the effect with $M=0$. }
\end{center}
\end{figure}
To see the nature of the horizon closely and the thermodynamical
characteristics of this gup-inspired black hole  we rewrite the
Lapse function in the following form.
\begin{equation}
f(r)=1- \frac{\frac{2G_0Mr^2}{c^2}}{r^3+
\frac{2G_0M\eta^2\hbar^2}{c^2}}
 =\frac{{\cal K}(r;r_{sr},\eta)}{r^3+
\tilde{\eta}^2r_{sr}} \label{lapse}
\end{equation}
\begin{equation}
{\cal K}(r;r_{sr},\eta)=r^3-r_{sr}r^2+ \tilde{\eta}^2r_{sr}
\label{poly1}
\end{equation}
Where $r_{sr}=\frac{2G_{0}M}{c^2}$, $M$ is the gravitational
energy-mass of the system and $\tilde{\eta}= \eta\hbar^2$,  The
lapse function can be expand as follows
\begin{equation}
f(r)\simeq 1-\frac{2G_{0}M}{c^2r}+\frac{4G^2_0M^2\tilde{
\eta}^2}{rc^4}-..... \label{f10}
\end{equation}
which is asymptotically flat. But, when $r << \eta$, it becomes
similar to that of de Sitter i.e.,
\begin{equation}
f(r)\simeq
1-\frac{r^2}{\tilde{\eta}^2}+\frac{c^2r^5}{2G_0M\tilde{\eta}^4}-......
\label{f20}
\end{equation}
So it is is flat and regular at $r=0$.

And, accordingly Eqn. (\ref{f1}) and (\ref{f2}) can be expressed
in terms of $\eta$. In natural unit $\hbar=1$. With this
approximation  the possible horizons is obtain seating $(\cal K
=0)$. Eqn. (\ref{poly1}) is containing the polynomial expression
of $(\cal K) $. To have a solution using Carnado's method we
define $r=z+\frac{2G_{0}M}{3c^{2}}=z+\frac{r_{sr}}{3}$. Eqn.
(\ref{poly1}) can be written in its canonical form as
\begin{equation}
{\cal K}(r;r_{sr},\eta)\equiv
z^3-(\frac{\xi_{1}}{4})z-(\frac{\xi_{2}}{4}) \label{poly2}
\end{equation}
where $\xi_{1}$ and $xi_{2}$ are the Weierstrab invariants given
by
\begin{equation}
\xi_{1}=\frac{4r_{sr}^2}{3}=\frac{4(2G_{0}M)^{2}}{3c^4}>0
\end{equation}
\begin{equation}
\xi_{2}=\frac{8r_{sr}^3}{27}\Big[1-\frac{\hbar^2\eta^2}{\tilde{\eta}^{2}}\Big]=
\frac{8(2G_{0}M)^{3}}{27c^6}\Big[1-\frac{\tilde{\eta}^2}{\tilde{\eta}_0^2}
\Big]
\end{equation}
And $\tilde{\eta}_0$ was conveniently defined as
\begin{equation}
\tilde{\eta}_0\equiv
\sqrt{\frac{2}{27}}r_{sr}=\sqrt{\frac{2}{27}}\frac{(2G_{0}M)}{c^2}
\end{equation}
Accordingly, the cubic discriminant associated with the polynomial
(\ref{poly2}) becomes
\begin{equation}
\Delta_{c}=
\frac{g_{2}^{3}}{16}\Big[1-\Big(1-\frac{\tilde{\eta}^2}{\tilde{\eta}_{0}^{2}}\Big)^2\Big]
\end{equation}
which vanishes trivially at
\begin{equation}
\tilde{\eta}= \tilde{\eta}_1=0
\end{equation}
and at a nontrivial $\tilde{\eta}$ at
\begin{equation}
\tilde{\eta} =\tilde{\eta}_2=
\frac{2}{\sqrt{27}}\frac{(2G_{0}M)}{c^2}
\end{equation}
The sign of the discriminant  $\Delta_{c}$ determines the the
nature of the roots. It is a  cubic equation. It is known to have
following possibilities of having its root. For $\Delta_{c} > 0$
when $\tilde{\eta}$ lies within the range $(\tilde{\eta}_{1}
<\tilde{\eta} <\tilde{\eta}_ {2})$, a multiple root for
$\Delta_{c}=0 (\tilde{\eta}=\tilde{\eta}_{1}$ or
 $\tilde{\eta}=\tilde{\eta}_{2})$, and a pair of complex root along with  a real root
 $\Delta_{c}<0 (\tilde{\eta}_{2}<\tilde{\eta}<\infty)$.

When $\tilde{\eta}$ lies in the  domain
$\tilde{\eta}_{1}<\tilde{\eta}<\tilde{\eta}_{0}$, the fundamental
Eqn. (\ref{poly2}) retains its standard form
\begin{equation}
4z^3-\xi_{1}\xi z-\xi_{2}=0, \label{XI}
\end{equation}
Here we have the scope  compared equation (\ref{XI}) with the
trigonometric identity
\begin{equation}
4cos^{3}\gamma-3cos\gamma-cos3\gamma=0,
\end{equation}
and obtained roots by applying some standard methods and recalling
$r=z+\frac{r_{sr}}{3}=z+\sqrt{\frac{\xi_{1}}{12}}$
\begin{equation}
{\cal R}_{+}\equiv \frac{r_{+}}{r_{sr}}=\frac{1+2cos\gamma}{3}
\label{root1}
\end{equation}
\begin{equation}
{\cal R}_{-}\equiv
\frac{r_{-}}{r_{sr}}=\frac{1-cos\gamma}{3}+\frac{\sqrt{3}}{3}sin\gamma
\end{equation}
\begin{equation}
{\cal R}_{n}\equiv
\frac{r_{n}}{r_{s}}=\frac{1-cos\gamma}{3}-\frac{\sqrt{3}}{3}sin\gamma
\end{equation}
for $\tilde{\eta}_{1}\leq \tilde{\eta} \leq \tilde{\eta}_{0}$,
where
\begin{equation}
\gamma=\frac{1}{3}cos^{-1}\Big(1-\frac{\tilde{\eta}^2}{\tilde{\eta}_{0}^{2}}\Big)
=\frac{1}{3}cos^{-1}\Big(1-\frac{27 \tilde{\eta}
c^{4}}{8G^2_0M^2}\Big) \label{root2}
\end{equation}

However, when $\tilde{\eta}$ lies in the domain
$\tilde{\eta}_{0}\leq l \leq \tilde{\eta}_{2}$, Eqn. (\ref{poly2})
can be written down as
\begin{equation}
4z^3-xi_{1}z+|\xi_{2}|=0.
\end{equation}
in this case we exploit the known trigonometric identity to find
the solution
\begin{equation}
4sin^{3}\gamma-3sin\gamma+sin3\gamma=0
\end{equation}
Like the former one, we obtained three real roots which are given
by
\begin{equation}
{\cal R}_{+}\equiv
\frac{r_{+}}{r_{sr}}=\frac{1-sin\gamma}{3}+\frac{\sqrt{3}}{3}cos\gamma
\label{root3}
\end{equation}
\begin{equation}
{\cal R}_{-}\equiv \frac{r_{-}}{r_{s}}=\frac{1+2sin\gamma}{3}
\end{equation}
\begin{equation}
{\cal R}_{n}\equiv
\frac{r_{n}}{r_{s}}=\frac{1-sin\gamma}{3}-\frac{\sqrt{3}}{3}cos\gamma
\end{equation}
where
\begin{equation}
\gamma=\frac{1}{3}sin^{-1}\Big(\frac{\tilde{\eta}^2}{\tilde{\eta}_{0}^{2}}
-1\Big)=\frac{1}{3}sin^{-1}\Big(\frac{27 \tilde{\eta}
c^4}{8G^2_0M^2}-1\Big)  \label{root4}
\end{equation}
Let us now see some general properties related to obtained roots.
If ${\cal R}_{+}(\gamma)$ is the function given by Eqns.
(\ref{root1}) or (\ref{root3}) with $\gamma=\gamma (r_{sr},l)$
given by Eqns. (\ref{root2}) or (\ref{root4}), then in the domain
$l_{1}\leq l \leq l_{2}$ we obtain the following
\begin{equation}
\Big(\frac{d{\cal R}_{+}}{d\gamma}\Big)=A(\gamma) \label{par1}
\end{equation}
\begin{equation}
r_{s}\Big(\frac{\partial\gamma}{\partial
r_{sr}}\Big)_{l}=-l\Big(\frac{\partial\gamma}{\partial
l}\Big)_{r_{sr}}=B(\gamma)  \label{par2}
\end{equation}
\begin{equation}
r_{s}\Big(\frac{\partial {\cal R}_{+}}{\partial
r_{s}}\Big)_{l}=-l\Big(\frac{\partial {\cal R}_{+}}{\partial
l}\Big)_{r_{s}}=C(\gamma)  \label{par3}
\end{equation}
where
\begin{equation}
 C(\gamma) =A(\gamma)B(\gamma)
 \end{equation}
In the domain $\tilde{\eta}_{1}\leq \tilde{\eta} \leq
 \tilde{\eta}_{0}$, the functions $ A(\gamma), B(\gamma)$ and $
C(\gamma)$ can be expressed as
\begin{equation}
A(\gamma)=-\frac{2}{3}sin\gamma,
\end{equation}
\begin{equation}
B(\gamma)=-\frac{2}{3}(cosec3\gamma-cot3\gamma),
\end{equation}
\begin{equation}
C(\gamma)=\frac{4}{9}sin\gamma(cosec3\gamma-cot3\gamma).
\end{equation}
However, in the domain $\tilde{\eta}_{0}\leq \tilde{\eta} \leq
\tilde{\eta}_{2}$ the functions $A(\gamma), B(\gamma)$ and $
C(\gamma)$, are given by
\begin{equation}
A(\gamma)=-\frac{1}{3}(cos\gamma+\sqrt{3}sin\gamma)
\end{equation}
\begin{equation}
B(\gamma)=-\frac{2}{3}(sec3\gamma+tan3\gamma)
\end{equation}
\begin{equation}
C(\gamma)=\frac{2}{9}(cos\gamma+\sqrt{3}sin\gamma)(sec3\gamma+tan3\gamma)
\end{equation}
This all bout the properties related to the nature of the roots of
the lapse function of this GUP inspired  black hole. Now we are in
a position to study the impact of GUP parameter on the
thermodynamical characteristics of this black hole to which we now
turn in the following section
\section{Thermodynamics of the GUP inspired quantum gravity corrected black hole}
We consider the Bekenstein-Hawking approach to study the
thermodynamics of this GUP inspired black hole \cite{BEK0, BEK1,
BEK2, BEK3, HAWKING}. It is necessary to keep in mind that through
this procedure, the construction of the thermodynamics is
implicitly established in the Carath Aodory framework, in which,
the existence of a function, termed as metrical entropy, is
ensured along with the absolute temperature. The  approach is
based on the  on the following postulate that  the event horizon
area of a black hole cannot decrease. It increases in most of the
transformations of the black hole \cite{MISNER}.  There fore
following using the Bekenstein  approach we consider $A=4\pi
r_{+}^{2}$ to be the area of the event horizon. Therefore the
entropy is then given by
\begin{equation}
S=\frac{k_{B}}{4}\frac{4\pi r_{+}^{2}}{l_{p}^{2}}
\end{equation}
where $k_{B}$ is the Boltzmann constant and $l_{p}$ is the Planck
length. It is therefore useful to define the entropy function as
\begin{equation}
S(r_{sr},\tilde{\eta})\equiv \frac{4l_p^2S}{4\pi k_B}
=r_{sr}^{2}{\cal R}_{+}^{2}(r_{sr},\tilde{\eta}) \label{entropy}
\end{equation}
\begin{equation}
\Big(\frac{\partial S}{\partial r_{sr}}\Big)dr_{sr}
+\Big(\frac{\partial S}{\partial \tilde{\eta}}\Big) d\tilde{\eta}
\end{equation}
which allows us to define the temperature parameter $T$ as
\begin{equation}
\frac{1}{T}=\Big(\frac{\partial S}{\partial
r_{sr}}\Big)_{\tilde{\eta}}
\end{equation}
solving it using Eqns. (\ref{par1}), (\ref{par3}) we have
\begin{equation}
T=\frac{1/(2r_{sr})}{{\cal R}_{+}\Big({\cal
R}_{+}+r_{sr}\frac{\partial {\cal R}_{+}}{\partial r_{s}}\Big)}
\end{equation}
\begin{equation}
T=\frac{T_{s}}{{\cal R}_{+}({\cal R}_{+}+ C{\gamma})} \label{T1}
\end{equation}
where $T_{s}=\frac{1}{2r_{sr}}$ is the temperature corresponding
to
 the Schwarzschild black hole.
For small values of $\tilde{\eta}$, $(\tilde{\eta} << r_{sr})$,
the temperature remains near $T_{s}$ and under this case $T$ can
be expanded as
\begin{equation}
T\simeq
T_{s}\Big(1-\frac{3\tilde{\eta}^4}{r_{sr}^{4}}-\frac{20\tilde{\eta}^6}{r_{sr}^{6}}-............\Big)
\end{equation}
\begin{equation}
T\simeq T_{s}\Big(1-\frac{3\eta^{2}\hbar^{4}c^{8}}{(2G_{0}M)^4}
-\frac{20\eta^{3}\hbar^{6}c^{12}}{(2G_{0}M)^6}-\frac{117\eta^{4}\hbar^{8}c^{16}}{(2G_{0}M)^8}-.....\Big)
\end{equation}
\begin{figure}[H]
 \begin{center}
\includegraphics[scale=0.5]{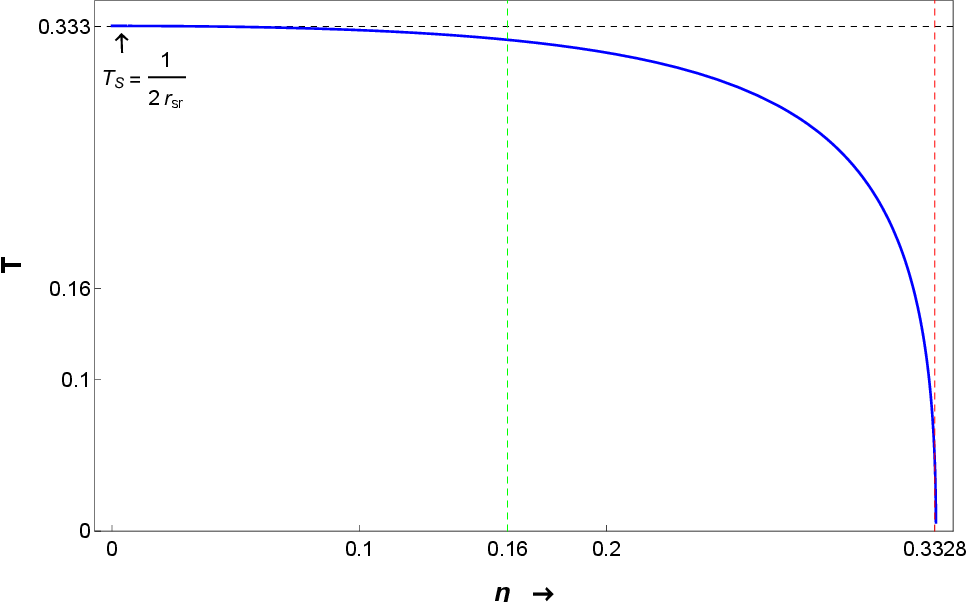} \caption{Variation of Temperature  parameter $T$ as a function of $\eta$. It has maximum value $T_{s}=\frac{1}{2r_{sr}}$ for values of $\eta < 0.16$ and minimum value $0$ at $\eta =0.3328$.}
 \end{center}
\end{figure}

If we  then generalized mechanical force is now defined by
\begin{equation}
F_{\tilde{\eta}}= -T\Big(\frac{\partial S}{\partial
\tilde{\eta}}\Big)_{r_{sr}}=\frac{r_{sr}}{\eta}\frac{C(\gamma)}{\big({\cal
R}_{+}(\gamma)+ C(\gamma)\big)} \label{force1}
\end{equation}
Introduction of the conjugated canonical pair $(\eta,
F_{tilde{\eta}})$ allow us to define  free energy ($F_{e}$)
through the following relation
\begin{equation}
F_{e}\equiv F_{\eta}\eta=\frac{C(\gamma)}{\big({\cal
R}_{+}(\gamma)+ C(\gamma)\big)}r_{sr}
\end{equation}

Note that when  $\tilde{\eta} << r_{sr}$, the free energy behaves
as
\begin{equation}
F_{e}\simeq
\frac{2\tilde{\eta}^2}{r_{sr}}\Big(1+\frac{3\tilde{\eta}^2}{r_{sr}^{2}}+\frac{12\tilde{\eta}^4}{r_{sr}^{4}}+.....\Big)
\end{equation}
\begin{equation}
F_{e}\simeq \frac{2\eta \hbar^{2}c^2}{(2G_{0}M)}\Big(1+\frac{3\eta
\hbar^{2}c^4}{(2G_{0}M)^{2}}+\frac{12\eta^{2}
\hbar^{4}c^8}{(2G_{0}M)^{4}}+......\Big)
\end{equation}
\begin{figure}[H]
 \begin{center}
\includegraphics[scale=0.5]{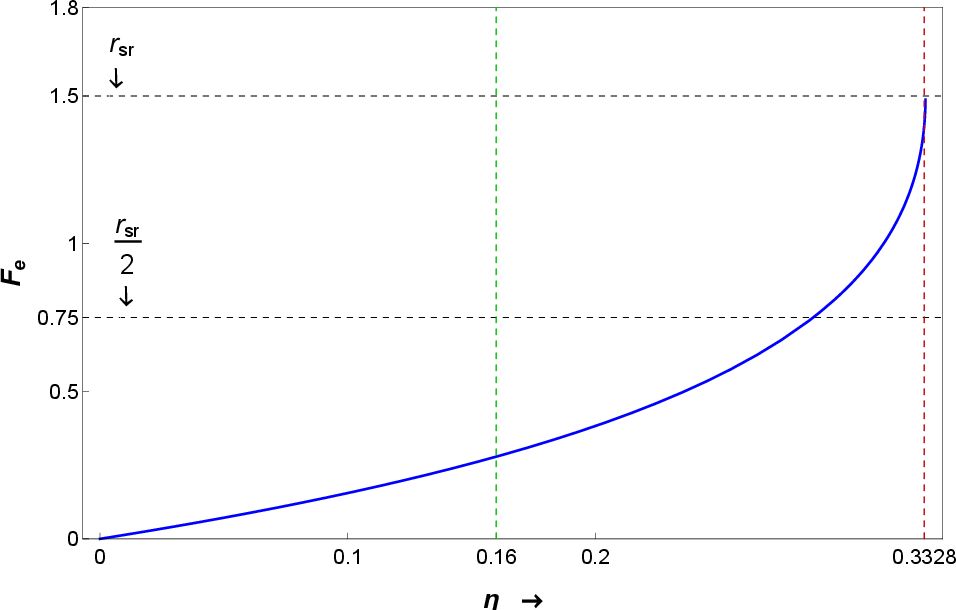}
 \caption{Variation of free energy parameter $F_{e}$ as a function of $\eta$. It has minimum value $0$ at $\eta=0$ and maximum value $r_{sr}$ at $\eta =0.3328$.}
 \end{center}
\end{figure}

If we combining the first and second laws of thermodynamics it
enables us to we can write \cite{LRE}
\begin{equation}
TdS\geq dr_{s}-F_{\eta}d\eta      \label{TDS1}
\end{equation}
From Eqn. (\ref{entropy}), entropy is a homogeneous function of
the second order in $r_{s}$ and $l$. Accordingly Euler's theorem
requires
\begin{equation}
2TS=r_{sr}-F_{\eta}\eta=r_{sr}-F_e
\end{equation}
This is the fundamental equation or the Gibbs-Duhem relation for
this GUP-inspired  black hole. Also, the equation of state (EoS)
relating the thermal state variable, $T$ , to the mechanical
variables of the system, can be obtained by combining Eqns.
(\ref{T1}) and (\ref{force1}):
\begin{equation}
\frac{F_{\tilde{\eta}}}{T}=\frac{2r_{sr}R_{+}(\gamma)C(\gamma)}{\tilde{\eta}}
\end{equation}
If the mass-energy $r_{sr}$ can treat as free energy in the sense
that the work can be stored in the form of potential energy and
that be recovered later. In fact, the total mass-energy
differential can be rewritten from Eqn. (\ref{TDS1}) as
\begin{equation}
dr_{s}\leq TdS + F_{\tilde{\eta}}d\tilde{\eta}
\end{equation}
where the inequality holds for spontaneous changes. Hence, for
reversible changes,
\begin{equation}
T=\Big(\frac{\partial r_{sr}}{\partial S}\Big)_{\tilde{\eta}}
\end{equation}
\begin{equation}
F_{\tilde{\eta}}=\Big(\frac{\partial r_{sr}}{\partial l}\Big)_{S}
\end{equation}
From these,one of the Maxwell relations is obtained as
\begin{equation}
\Big(\frac{\partial T}{\partial
\tilde{\eta}}\Big)_{S}=\Big(\frac{\partial
F_{\tilde{\eta}}}{\partial S}\Big)_{\tilde{\eta}}
\end{equation}
If we consider the black hole as a closed and isolated system, we
can write
\begin{equation}
\Delta r_{s}=\Delta Q -\delta W
\end{equation}
for changes in the black hole's  mass-energy, where $\Delta Q$ is
the heat flux across the surface of the black hole, and $\Delta W$
is the work. In a reversible process, with $S$ and $\tilde{\eta}$
as constants, we can now express the corresponding enthalpy as
\begin{equation}
H\equiv r_{sr}-F_{\tilde{\eta}}\eta = 2TS
\end{equation}
which is obtained by adding an additional energy, that accounts
for a mechanical coupling. Using Eqns. (\ref{par1})-(\ref{par3}),
we get
\begin{equation}
H=\frac{2G_{0}M}{c^2}\frac{\cal{R}_{+}}{({\cal R}_{+}(\gamma)+
C{\gamma})}
\end{equation}
when  $\tilde{\eta} << r_{sr}$, the enthalpy behaves as
\begin{equation}
H\simeq
r_{sr}\Big(1-\frac{2\tilde{\eta}}{r_{sr}^{2}}-\frac{6\tilde{\eta}^2}{r_{sr}^{4}}-\frac{24\tilde{\eta}^3}{r_{sr}^{6}}-........\Big)
\end{equation}
\begin{equation}
H\simeq \frac{(2G_{0}M)}{c^2}\Big(1-\frac{2\eta
\hbar^{2}c^4}{(2G_{0}M)^{2}}-\frac{6\eta^{2}
\hbar^{4}c^8}{(2G_{0}M)^{4}}-........\Big)
\end{equation}

\begin{figure}[H]
 \begin{center}
\includegraphics[scale=0.5]{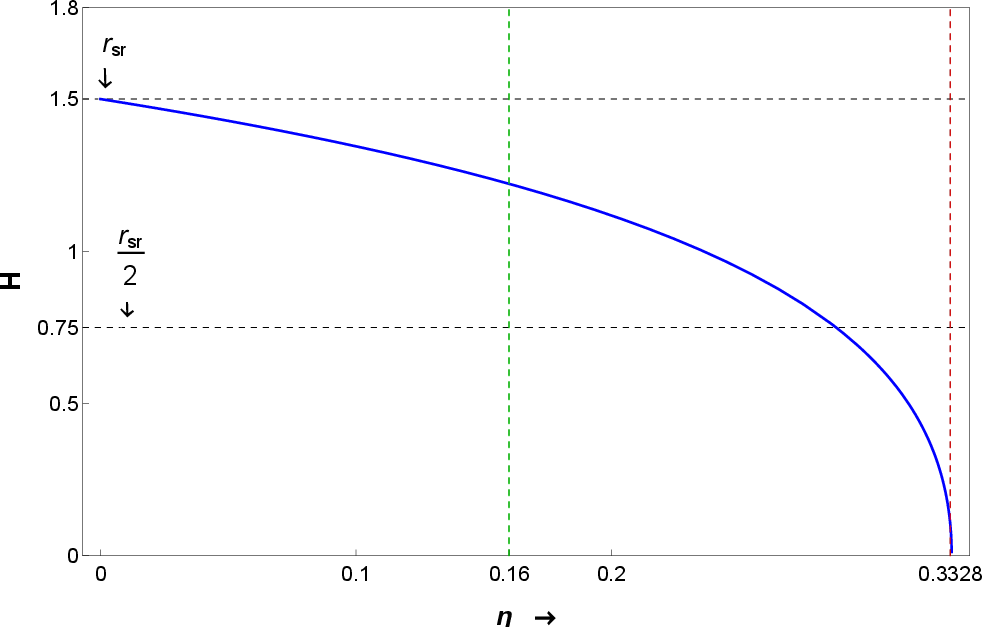} \caption{Variation of Enthalpy parameter $H$ as a function of $\eta$. It has minimum value $0$ at $\eta=0.3328$ and maximum value $r_{sr}$ at $\eta =0$.}
 \end{center}
\end{figure}

Finally, by adding thermal coupling term, we can express the
Helmholtz free energy in terms of $r_{sr}$ in the following form
\begin{equation}
F_{H}=r_{sr}-TS=TS+\Xi
\end{equation}
Using equations (\ref{par1}) - (\ref{par3}), The above expression
becomes
\begin{equation}
F_{H}= \frac{r_{sr}}{2}\Big[\frac{{\cal
R}_{+}(\gamma)+2C(\gamma)}{{\cal R}_{+}(\gamma)+C(\gamma)}\Big]
\end{equation}
\begin{figure}[H]
 \begin{center}
\includegraphics[scale=0.5]{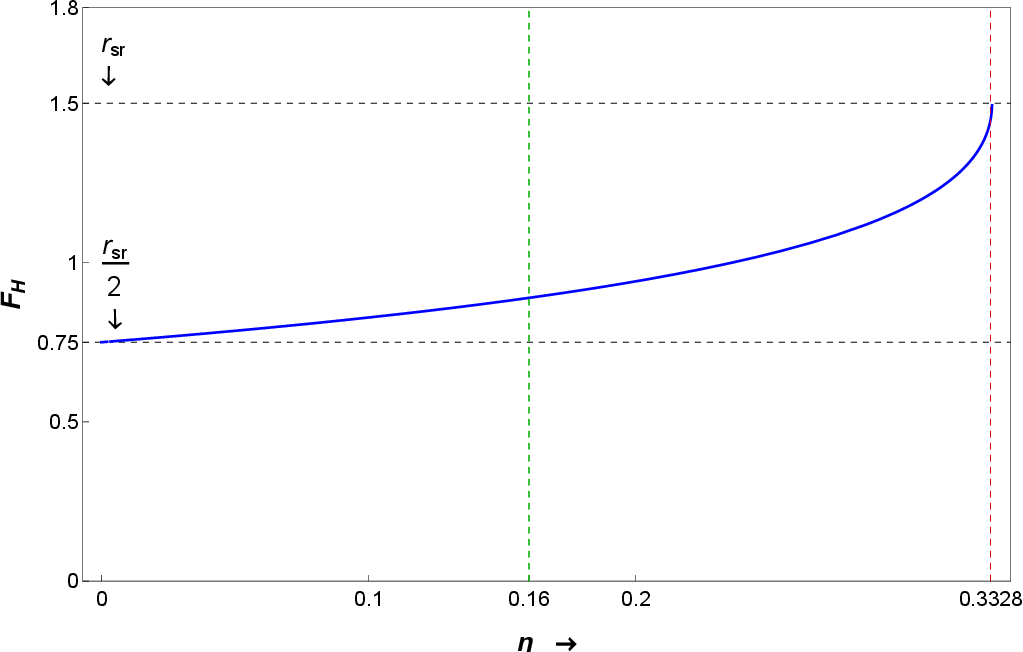} \caption{Variation of Helmholtz free energy parameter $F_{H}$ as a function of $\eta$. It has minimum value $\frac{r_{sr}}{2}$ at $\eta=0.16$ and maximum value $r_{sr}$ at $\eta =0.3328$.}
 \end{center}
\end{figure}
This all about the thermodynamics of this GUP inspired Black hole.
In the following section we will study the impact of GUP parameter
on the accretion process on to the black hole associated with the
GUP inspired quantum corrected spacetime.
\section{Description of the geodesic in the GUP inspired quantum
corrected spacetime geometry}
In this article we are intended to
study   the accretion phenomena onto a spherically symmetric
Schwarzschild black hole employing a modified uncertainty relation
that admits a quantum gravity correction that finds its place
holding the hand of the concept of minimal measurable length.
Although, other models which are associated with the GUP
correspond to the concept of minimum measurable length and maximum
measurable momentum length simultaneously. This type of problem is
amenable we to have an answer  for all kinds  of available
generalization of uncertainty relation \cite{MEG1, MEG2, ALI1,
ALI2, PED, ARN, HOMA, SG, SG1, ANI, ANI1}. We will only consider
the generalization related to the existence of a minimal length.
During this context we consider steady, accretion onto a modified
static and spherically symmetric Schwarzschild black hole. We
obtain the critical point, critical fluid velocity, temperature,
mass accretion rate, and observed total integrated flux within the
proposed GUP framework.

The line element in Eqn. (\ref{SCH1}) can be written down within
the form of a Schwarzschild-like metric by introducing an
effective mass $M_{eff}(r)$ which could be a  function of the
radial coordinate and depends on the free parameter $\tilde{\xi}$,
that is
\begin{equation}
ds^2=-\Big(1-2\frac{M_{eff}(r)}{r}\Big)dt^2
+\Big(1-2\frac{M_{eff}(r)}{r}\Big)^{-1}dr^2+r^2(d\theta^2
+sin^2\theta d\phi^2)
\end{equation}
with
\begin{equation}
M_{eff}(r)=\frac{M}{(1+\frac{2M\eta}{r^3})} \label{Meff1}
\end{equation}
Putting $\tilde{\eta}=\frac{M\eta}{M^3}=\frac{\eta}{M^2}$ and
$x=\frac{r}{M}$, Eqns.(\ref{Meff1}) can be rewritten as
\begin{equation}
M_{eff}(r)=\frac{M}{(1+\frac{2\tilde{\eta}}{x^3})} \label{MEFF}
\end{equation}
Plots of the effective mass per unit mass as a function of $x$ is
shown below.
\begin{figure}[H]
\begin{center}
\includegraphics[scale=0.5]{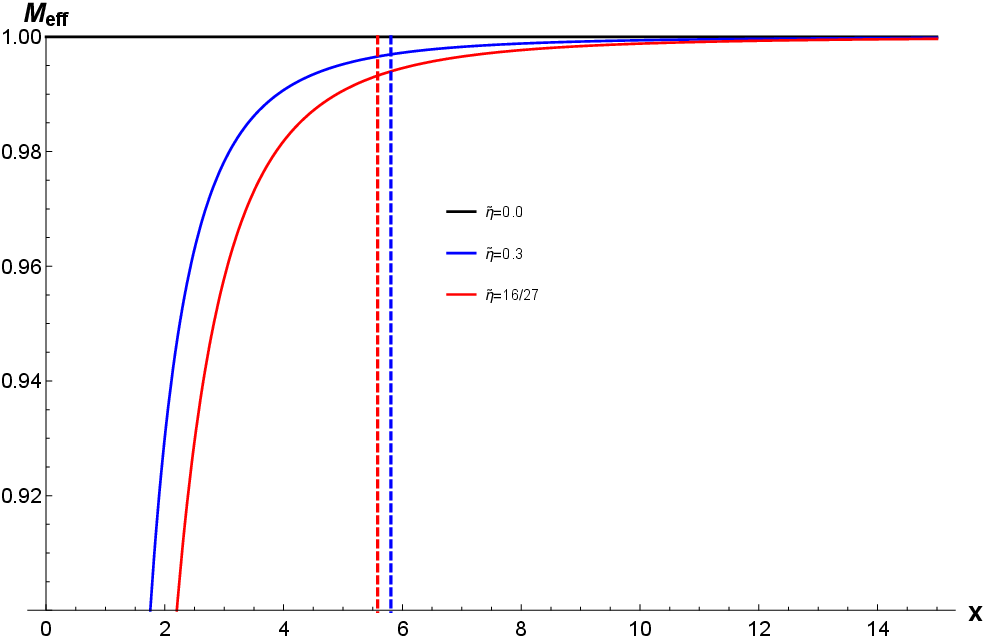}
\caption{Plots of the effective mass as a function of $x$ for
$\eta= \tilde{\eta}_c= 16/27$ (red) and $\tilde{\eta}=0.3$ (blue).
The vertical dashed lines indicate the ISCO for the same values of
$\tilde{\eta}$ respectively}. \label{figure}
\end{center}
\end{figure}
Using this definition (\ref{MEFF}) we can write write down the
Lagrangian from the modified metric endowed with quantum
correction replacing $M$ by $M \rightarrow M_{eff}(r)$.
\begin{equation}
{\cal L}=\frac{1}{2}[-{\cal F}(r)\dot{t}^2 +{-1}\frac{1}{{\cal
F}(r)}\dot{r}^2+r^2\dot{\phi}^{2}] \label{LAG}
\end{equation}
where $\Big(1-2\frac{M_{eff}(r)}{r}\Big)={\cal F}(r)$ and the over
dots referees to  derivative with respect to the affine parameter.
We have restricted ourselves to the equatorial plane setting
$(\theta=\pi/2,\dot{\theta}=0)$. This indeed does not loose any
generality. From Eqn.(\ref{LAG}), the generalized momenta are are
computed as follows
\begin{equation}
p_{t}=\frac{\partial{\cal L}} {\partial\dot{t}}=
-\Big(1-2\frac{M_{eff}(r)}{r}\Big)\dot{t}=-{\cal K}, \label{mom1}
\end{equation}
\begin{equation}
p_{r}=\frac{\partial{\cal L}}
{\partial\dot{r}}=\Big(1-2\frac{M_{eff}(r)}{r}\Big)^{-1}\dot{r}=
{\cal B} \label{mom2}
\end{equation}
\begin{equation}
p_{\phi}=\frac{\partial{\cal L}}
{\partial\dot{\phi}}=r^2\dot{\phi} = {\cal A } \label{mom3}
\end{equation}
The constants ${\cal K}$ and $ {\cal A }$ are representing the
energy and angular momentum per unit rest mass of the particle
respectively,  The canonical Hamiltonian is abstained by the
Legendre transformation
\begin{equation}
{\cal H}=p_{t}\dot{t}+p_{r}\dot{r}+p_{\phi}\dot{\phi}-{\cal L}
\end{equation}
and it reads
\begin{equation}
{\cal H}=-{\cal K}\dot{t}+{\cal B}\dot{r}+{\cal A
}\dot{\phi}=-\frac{1}{2} \label{hamil}
\end{equation}
Note that it is independencies of time so is a constant quantity.
The particle is assumed at rest at infinity that allows to set the
righthand side of the equation equal to $-1$. We obtain the the
equation corresponding to the energy  plugging in
Eqns.(\ref{mom1}) and (\ref{mom3})in to the  Eqn.(\ref{hamil})
\begin{equation}
\frac{1}{2}\Big(\frac{dr}{dt}\Big)^2+ V_{eff}(r)=\frac{1}{2}({\cal
K}^2-1), \label{energy1}
\end{equation}
where $ V_{eff}(r)$ is given by
\begin{equation}
V_{eff}(r)=-\frac{M_{eff}(r)}{r}+\frac{h^2}{2r^2}-\frac{M_{eff}(r){\cal
A }^{2}}{r^3} \label{Veff}
\end{equation}
$V_{eff}(r)$ is nothing but the the effective potential per unit
mass in this situation. Let us now proceed to find out the
equation of the orbit of the massive particle of effective mass
$M_{eff}(r)$. To this end we define
\begin{equation}
\dot{r}=\frac{dr}{d\tau}
\end{equation}
By using Eqn.(\ref{mom3}) we can  write
\begin{equation}
\dot{r} =\frac{dr}{d\phi}\frac{d\phi}{d\tau}= \frac{{\cal A
}}{r^2}\frac{dr}{d\phi} \label{RDOT}
\end{equation}
It is beneficial to make the change of variable $u = 1/r$ at this
stage. In terms of $u$  Eqn.(\ref{energy1}) can be written down as
\begin{equation}
\Big(\frac{du}{d\phi}\Big)^{2}+u=\frac{({\cal K}^2-1)}{{\cal A
}^2}+\frac{2uM_{eff}(u)}{{\cal A }^2}+2u^{3}M_{eff}(u) \label{TRA}
\end{equation}
using Eqn (\ref{RDOT}).  Eqn. (\ref{TRA}) leads us to the
following equation of the orbit after undergoing the
differentiation with respect to $\phi$
\begin{equation}
\frac{d^{2}u}{d\phi^{2}}+u=\frac{M}{(1+2m\eta
u^3)}\Bigg[\frac{1}{{\cal A }^2}-\frac{6M\eta u^3}{{\cal A
}^{2}(1+2m\eta u^3)} +3u^2-\frac{6M\eta u^5}{(1+2m\eta u^3)}\Bigg]
\label{derivative2}
\end{equation}
\subsection{Condition of circular orbits of the  massive particles}
To determine the fundamental equations describing the
time-averaged radial disk structure, we first explicitly calculate
the  precise expression of angular momentum ${\cal A}$, the energy
${\cal K}$  per unit mass and the angular velocity $\omega$ of
particles taking possession of the circular trajectories.

For circular orbits within the  equatorial plane $\dot{r}= 0$.
Therefore that $r$ is constant, and consequently $u = 1/r$  may be
a constant indeed. We now define two dimensionless quantities
$x=r/M$ and $\tilde{\eta}=\frac{M\eta}{M^3}=\frac{\eta}{M^2}$.
From Eqn.(\ref{derivative2}) leads us we obtain an expression for
the specific angular momentum (angular momentum per unit mass)
${\cal A}$  and if we rewrite ${\cal A}$ in terms of the
dimensionless quantities $x$ and $\tilde{\eta}$ it reads
\begin{equation}
{\cal
A}=\frac{x^{2}M\sqrt{(x^3-4\tilde\eta)}}{\sqrt{(x^6-3x^5+4\tilde{\eta}x^3+4\tilde{\eta}^2)}}.
\label{angular}
\end{equation}
Putting $\bar{A}=\frac{{\cal A}}{M}$, variation of $\bar{A}$ with
$x$ is shown below.
\begin{figure}[H]
\begin{center}
\includegraphics[scale=0.5]{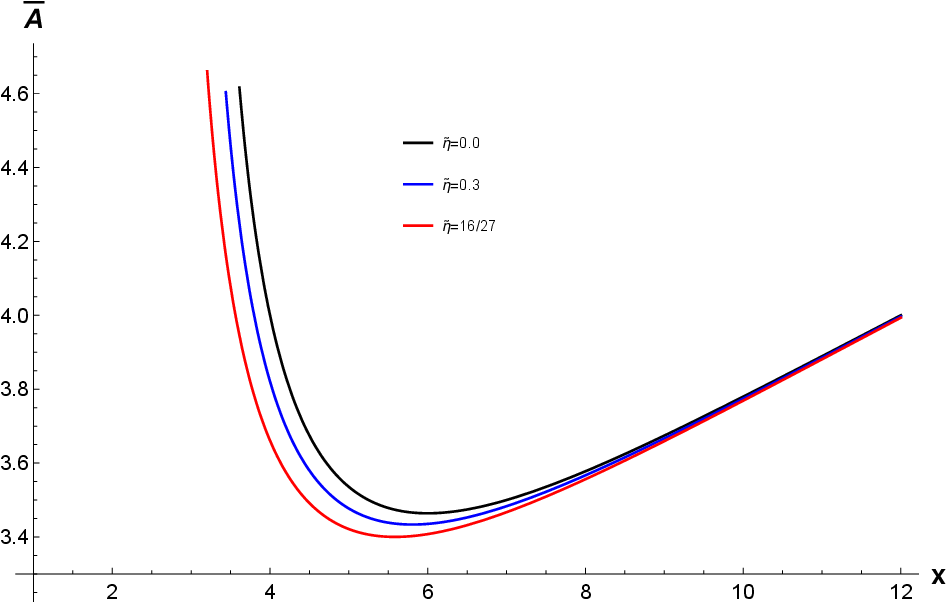}
\caption{The angular momentum $\bar{A} =\frac{{\cal A}}{M}$ vs.
$x$ for several values of $\tilde{\eta}$. From left to right:
$\tilde{\eta}=\tilde{\eta}_C=\frac{16}{27}$(red),
$\tilde{\eta}=0.3$(blue). The black solid curve corresponds to the
classical case $\tilde{\eta}=0$.}
\end{center}
\end{figure}
Making use of the condition of having circular orbit $\dot{r}= 0$
in Eqn.(\ref{energy1}) we can have the expression of the specific
energy (energy per unit mass) ${\cal K}$, and substituting the
expression for $\bar{ A}$ the specific energy ${\cal K}$ can be
expressed as
\begin{equation}
{\cal
K}=\frac{\sqrt{(x^3-2x^2+2\tilde{\eta})}\sqrt{(x^6-2x^5+4\tilde{\eta}x^3-4\tilde{\eta}x^2+
4\tilde{\eta}^2)}}{\sqrt{(x^3+2\tilde{\eta})}\sqrt{(x^6-3x^5+4\tilde{\eta}x^3+4\tilde{\eta}^2)}}.
\label{energy2}
\end{equation}
Upon substituting Eqn.(\ref{angular}) into Eqn.(\ref{Veff}), the
effective potential acquires the following form
\begin{equation}
V_{eff}=-\frac{x^2}{(x^3+2\tilde{\eta})}\Bigg[1-\frac{x^{2}(x^3-4\tilde{\eta})}{(x^6-3x^5+4\tilde{\eta}x^3+4\tilde{\eta}^2)}-
\frac{x^{2}(x^3-4\tilde{\eta})}{2(x^6-3x^5+4\tilde{\eta}x^3+4\tilde{\eta}^2)}\Bigg].
\end{equation}
Plot of $V_{eff}$ as function of $x$ for the same values of the
free parameter $\tilde{\eta}$ in the range $ 0\leq
\tilde{\eta}\leq 16/27$ is shown in Figure 4.
\begin{figure}[H]
 \begin{center}
\includegraphics[scale=0.5]{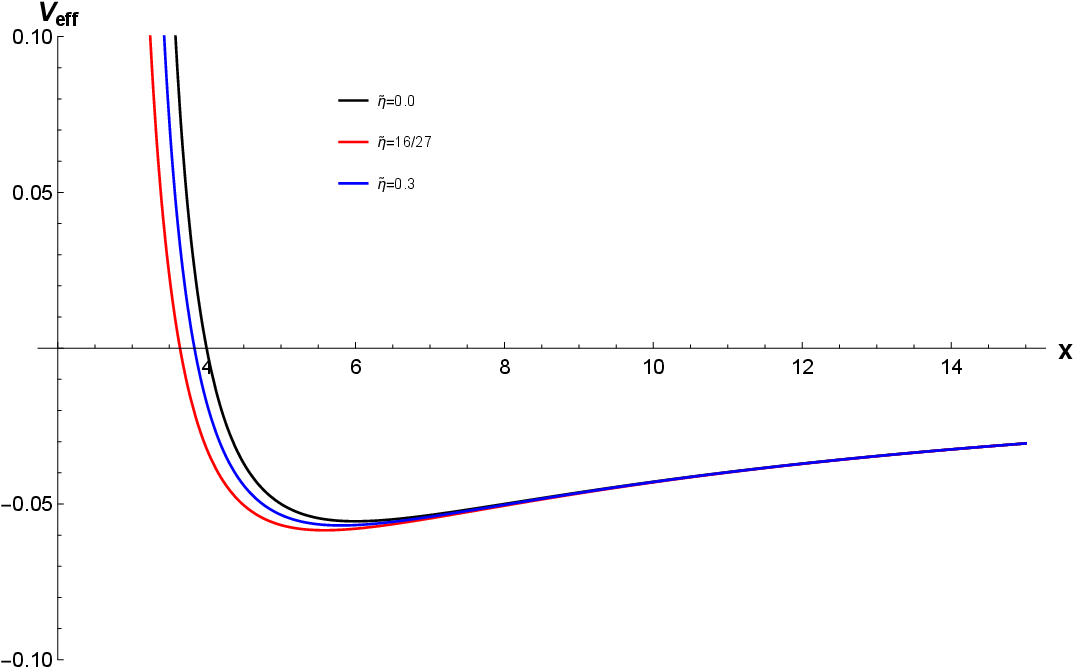}
\caption{Plot of the the effective potential for different values
of $\tilde{\eta}$. From left to right:
$\tilde{\eta}=\tilde{\eta}_C=\frac{16}{27}$(red),
$\tilde{\eta}=0.3$(blue). The black solid curve refers to the
classical case $\tilde{\eta}=0.$}
\end{center}
\end{figure}
Using Eqns. (\ref{angular}) and (\ref{energy2}) the angular
velocity can be computed from Eqns. (\ref{mom1}) and (\ref{mom3})
as
\begin{equation}
\omega=\frac{\dot{\phi}}{\dot{t}}=
\frac{\sqrt{(x^3-4\tilde{\eta})(x^3-2x^2+2\tilde{\eta})}}{M\sqrt{(x^3+2\tilde{\eta})
(x^6-2x^5+4\tilde{\eta}x^3-4\tilde{\eta}x^2+4\tilde{\eta}^2)}}.
\end{equation}
Note  that $V_{eff}$, $\bar{A}$, ${\cal K}$ and $\omega$ reduce to
the corresponding classical expressions in the limit
$\tilde{\eta}\to 0$.

At the local minima of the effective potential the possession of
circular orbits materialize. Thus, the  radius of the innermost
stable circular geodesic orbit $x_{isco}$ is found out from the
condition
\begin{equation}
 \frac{d^{2}V_{eff}}{dx^2}=0.
\end{equation}
Note that here it will appear in a dimension less manner. On the
other hand  ISCO can be calculated from  the following condition
\cite{PAGE}.
\begin{equation}
\frac{d{\cal A}}{dx}=\frac{d {\cal K}}{dx}=0.
\end{equation}
We are furnishing  the ISCO of the thin disk for a set of selected
values of $\tilde{\eta}$ in a tabular form
\begin{table}[H]
\begin{center}
\begin{tabular}{|c|cc|cc|cc|} \hline
    $\tilde{\eta}$             &&  $x_{isco}$    \\\hline
    $\frac{16}{27}$           &&  5.58396 \\
    $0.3$                     &&  5.80455  \\
    $0$                       &&   6  \\
 \hline
\end{tabular}
\caption{vales of $x_{isco}$ for different $\tilde{\eta}$}.
\end{center}
\end{table}
\begin{figure}[H]
 \begin{center}
\includegraphics[scale=0.5]{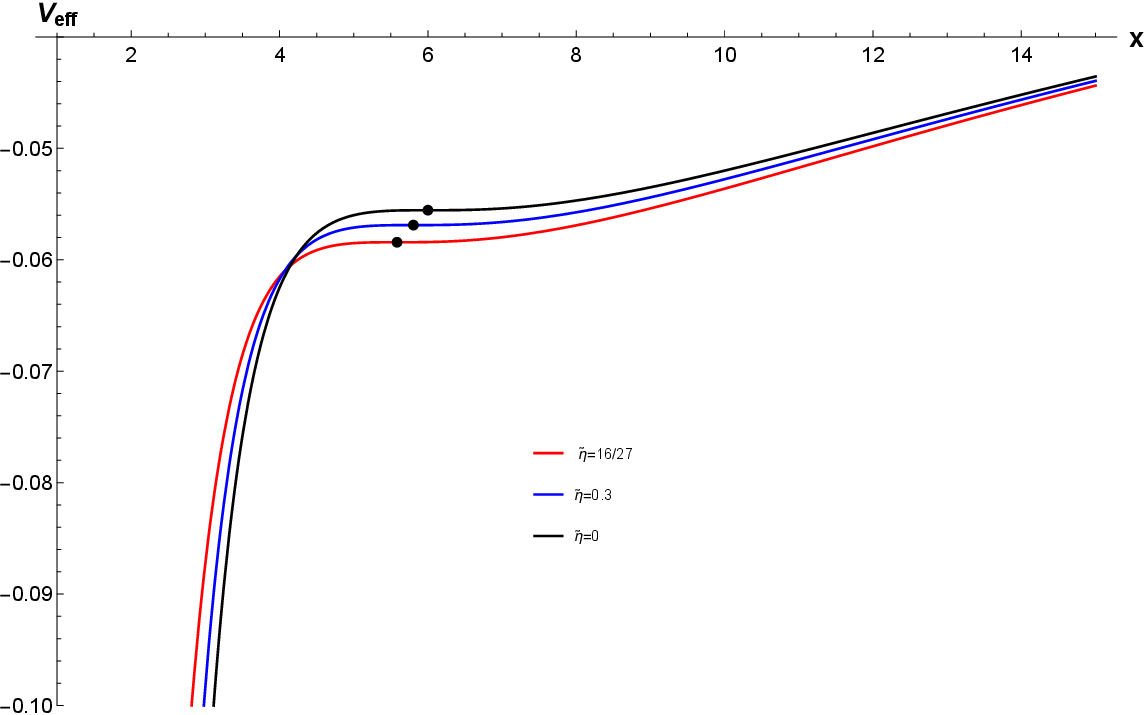} \caption{The sketch of effective potential
for the values of $\bar{A}$, evaluated at ISCO. From left to
right: $\tilde{\eta}=\tilde{\eta}_C=\frac{16}{27}$ (red),
$\tilde{\eta}=0.3$ (blue). The classical case is described by the
black solid curve where $\tilde{\eta}=0$.}
 \end{center}
\end{figure}
\begin{table}[H]
\begin{center}
\begin{tabular}{|c|cc|cc|cc|} \hline
    $\tilde{\eta}$     &&  ${\cal K}_{isco}$   && $\epsilon $  \\\hline
    $\frac{16}{27}$    &&  0.939752     &&  $6.0248$  \\
    $0.3$              &&  0.941393     &&  $5.8607$ \\
    $0$                &&  0.942809     &&  $5.7191$ \\
 \hline
\end{tabular}
\caption{The energy per unit mass at the ISCO and the efficiency
$\epsilon$ of the conversion of the accreted mass into radiation
for several values of $\tilde{\eta}$.}
 \end{center}
\end{table}
where, due to the fact that ${\cal A} $ is constant for circular
orbits, the derivatives must be calculated for $V_{eff}$ as given
by Eqn. (\ref{Veff}).
 This yields
\begin{eqnarray}
\frac{d^{2}V_{eff}}{dx^2}&=&-\frac{(x^6-28\tilde{\eta}x^3+8\tilde{\eta}^2)}{(x^3+2\tilde{\eta})^3}\nonumber\\
&+&\frac{3(x^3-
4\tilde{\eta})}{(x^6-3x^5+4\tilde{\eta}x^3+4\tilde{\eta}^2)}+\frac{6x^5(x^3-
4\tilde{\eta})}{(x^3+2\tilde{\eta})^{2}(x^6-3x^5+4\tilde{\eta}x^3+4\tilde{\eta}^2)}\nonumber\\
&-&\frac{18x^8(x^3-
4\tilde{\eta})}{(x^3+2\tilde{\eta})^{3}(x^6-3x^5+4\tilde{\eta}x^3+4\tilde{\eta})}.
\end{eqnarray}
\section{Mass accretion rate for thin accretion disk}
A general relativistic treatment of an accretion disk  around a
black hole was  reported  in the pioneering articles \cite{PAGE,
BAMBI}. Let us consider a simple non-relativistic model of an
accretion disk around a compact central object.  Recent
investigation on this issue in different perspective  are
\cite{HARKO, ROME, ARA, LUIS}. Here it is assumed that matter
spirals inwards by losing  angular momentum which, because of
turbulent viscous density be transferred outward through the disk.
As the gas moves inwards, it loses gravitational energy and heat
over the surroundings by emitting thermal radiation \cite{SHAKU}.
This model assumes that disk is in a quasi-steady state lying in
the equatorial plane of an stationary, axially-symmetric spacetime
back ground.  The disk material is assumed to be moving in nearly
geodesic circular orbit. The disk is so thin that its maximum
thickness D satisfies $D/2 R << 1$ where R refers to the
characteristic radius of the disk. The heat produced  by stress
and dynamical friction is efficiently emitted in the form of
radiation substantially from the surface of the disk. The amounts
describing the thermal characteristics of the disk  are averaged
over the azimuthal angle $\phi = 2\pi$, over the thickness $D$,
and over the time scale $\Delta \tau$. Here $\tau$ is the time
that the gas takes to flow inward through a distance $2D$. With
these propositions, the time-averaged radius of the disk  is
attained from the laws of conservation of rest mass energy and
angular momentum.  The integration of the equation of mass
conservation reveals that the mass accretion rate remains constant
for this process
\begin{equation}
\dot{M}= \frac{dM}{d\tau}= -2\pi\Sigma(r)v^r = constant,
\end{equation}
where $v^r$ and $\sigma$ are respectively the radial velocity and
surface density of the accretion disk. Now the combined criteria
of the energy and angular momentum conservation leads us  to find
out the  expression of the  differential of the luminosity
$dL_{\infty}$ at infinity \cite{PAGE, JOSHI}.
\begin{equation}
\frac{dL_{Inf}}{dlnr}=4\pi r\sqrt{-g}{\cal K}F(r), \label{Lumino}
\end{equation}
 where
$F$ is the the flux of radiant energy emitted from the upper face
of disk in the local frame of the accreting fluid. Let us call it
as $B(r)\equiv B{Mx}$ for later convenience. It has the expressed
in terms of the specific angular momentum ${\cal A}$, the specific
energy ${\cal K}$ and the angular velocity $\Omega$ which reads
\begin{equation}
F(r)=-\frac{\dot{M}}{4\pi \sqrt{-g}}\frac{1}{({\cal K}-\Omega
{\cal A})^2} \frac{d\Omega}{dr}\int_{r_{isco}}^{r}({\cal K}-\Omega
{\cal A})\frac{d{\cal A}}{dr}dr, \label{flux}
\end{equation}
where $\sqrt{-g}=r$ it remains the sane for  both for the
GUP-improved quantum corrected metric and for the classical
Schwarzschild metric. The numerical integration of Eqn.
(\ref{flux}) gets simplified by using the relation $\frac{d{\cal
K} }{dr}=\Omega\frac{d{\cal A}}{dr}$ \cite{PAGE} and integrating
by parts we have
\begin{equation}
\int_{r_{isco}}^{r}({\cal K}-\Omega {\cal A})\frac{d{\cal
A}}{dr}dr={\cal K}h-{\cal K}_{isco}{\cal A}_{isco}
-2\int_{r_{isco}}^{r}{\cal A}\frac{d{\cal A}}{dr}dr.
\end{equation}
Plot of energy flux per unit accretion rate $\frac{F(x)}{\dot{M}}$
from a thin accretion disk around a GUP-improved Schwarzschild
black hole as a function of $x$ is shown below.

\begin{figure}[H]
 \begin{center}
\includegraphics[scale=0.5]{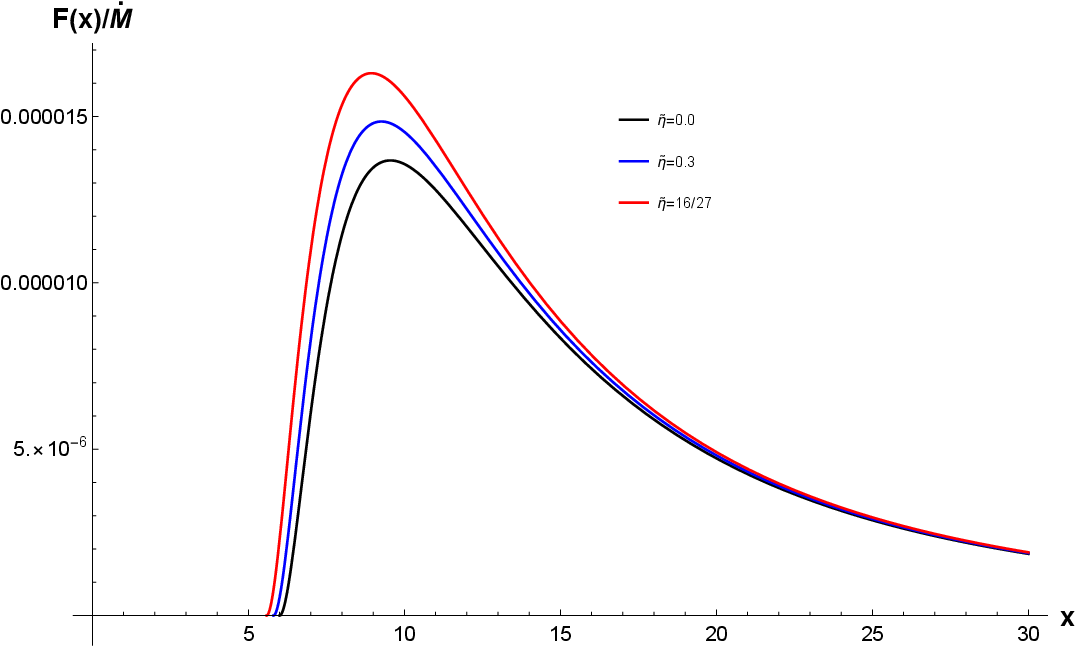} \caption{Energy flux per unit accretion rate
from a thin accretion disk around a GUP-improved Schwarzschild
black hole for $\tilde{\eta}=\tilde{\eta}_{c}=\frac{16}{27}$
(red),
 $\tilde{\eta}=0.3$ (blue).
The black solid curve is the energy flux from the disk around a
classical Schwarzschild black hole  $(\tilde{\eta}=0).$}
 \end{center}
\end{figure}
Plots of the differential luminosity at infinity per unit
accretion rate from a thin disk around a GUP-improved
Schwarzschild black hole as a function of $x$ is shown below.
\begin{figure}[H]
 \begin{center}
\includegraphics[scale=0.5]{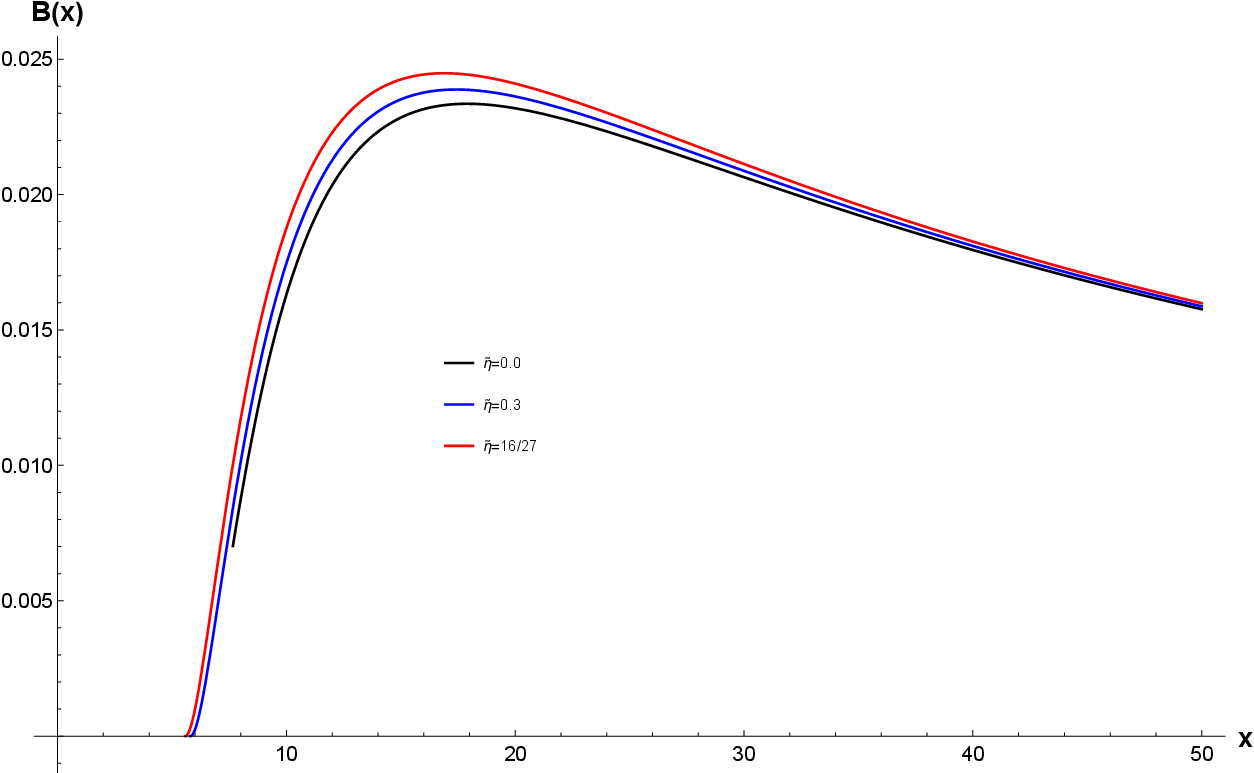}

\end{center}
\end{figure}
The radiation emitted can be considered as a black body radiation
with the temperature given by
\begin{equation}
T(r)=\sigma^{-\frac{1}{4}}F(r)^{\frac{1}{4}},
\end{equation}
as  it  is assumed that during the accretion process the disk in
thermodynamic equilibrium.  Here $\sigma$ stands  for
Stefan-Boltzmann constant.

The radial profile of the temperature of the accretion disk (more
precisely, the radial profile
$\frac{\sigma^{1/4}T}{\dot{M}^{1/4}}$), is shown for different
values of the dimensionless free parameter $\tilde{\eta}$ namely,
for the critical value $\tilde{\eta}_{c}=16/27$, for
$\tilde{\eta}=0.3$, and for the classical solution
$\tilde{\eta}=0$.

\begin{figure}[H]
 \begin{center}
\includegraphics[scale=0.5]{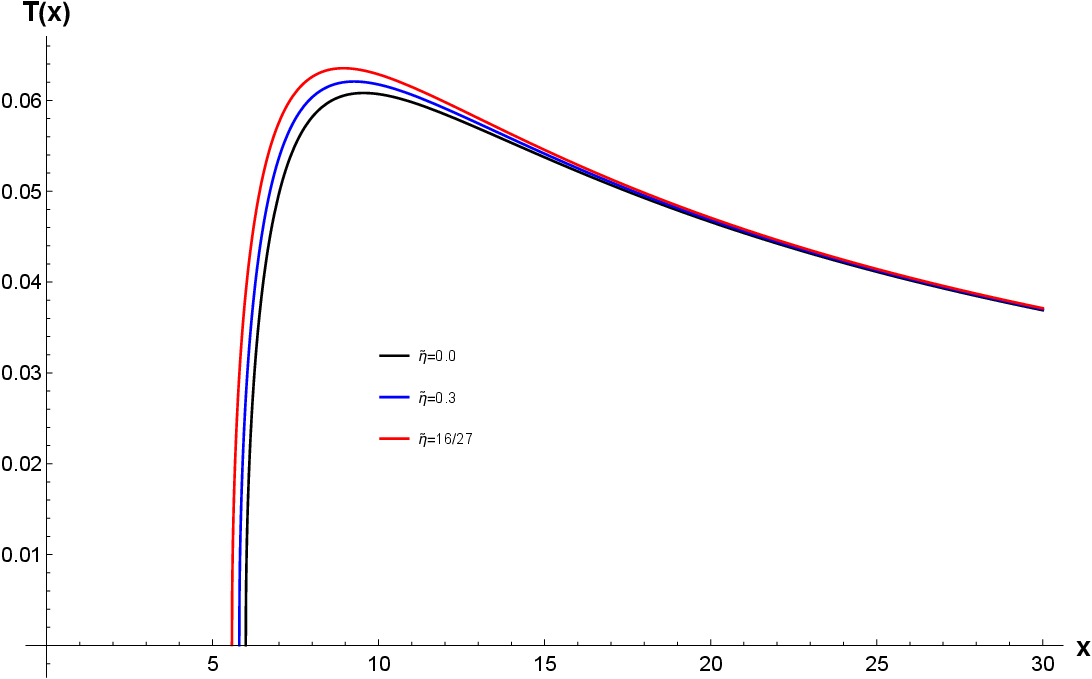} \caption{Radial profiles of the temperature
per unit accretion rate of a thin accretion disk around a
GUP-improved Schwarzschild black hole for
$\tilde{\eta}=\tilde{\eta}_{c}=\frac{16}{27}$ (red),
$\tilde{\eta}=0.3$ (blue). The black solid curve corresponds  to
the temperature of the disk around a Schwarzschild black hole in
general relativity
 $\tilde{\eta}=0.$}
 \end{center}
\caption{Differential luminosity at infinity per unit
accretion rate from a thin disk around a GUP-improved
Schwarzschild black hole for $\tilde{\eta}=\tilde{\eta}_{c}=\frac{16}{27}$ (red),
$\tilde{\eta}=0.3$ (blue). The black solid curve is the energy
flux from the disk around a classical Schwarzschild black hole
$(\tilde{\eta}=0).$}
\end{figure}

What follows next is a tabular presentation of furnishing maximum
value and increase in maximum value of  the time averaged energy
flux per unit accretion rate $\frac{F(x)}{\dot{M}}$, the
differential luminosity per unit accretion rate $B(x)$ and  the
radial profile of the temperature of the accretion disk $T(x)$ at
$\tilde{\eta}=$ 0,0.3, and $\frac{16}{27}$, respectively (
table-III) to observe at a glance the change of the peak values of
$\frac{F(x)}{\dot{M}}$, $B(x)$ and $T(x)$ in percentage  due to
the presence of quantum correction endowed through the GUP
framework.

\begin{table} [H]
\begin{center}
 \begin{tabular}{| c | c  p{1.8cm} | c  p{1.8cm} | c  p{1.8cm} | }
        \hline
        \multicolumn{1}{|c|}{$\tilde{\eta}$} & \multicolumn{2}{c|}{$\frac{F(x)}{\dot{M}}$}
         & \multicolumn{2}{c|}{$B(x)$} & \multicolumn{2}{c|}{$T(x)$}\\

     & Maximum value  & Increase in max value in $\%$ & Maximum value & Increase in max value in $\%$ & Maximum value & Increase in max value in $\%$ \\

  \hline

  0 & 0.00001367 & -- & 0.0233514 & -- & 0.0608144 & --\\
  0.3 & 0.00001484 & 8.55 & 0.0238780 & 2.25 & 0.0620756 & 2.07  \\  16/27 & 0.00001630 & 19.23 & 0.0244795 & 4.83 & 0.0635401 & 4.48 \\
    \hline
    \end{tabular}
\caption{Table containing maximum value and increase in maximum
value of  the time averaged energy flux per unit accretion rate
$\frac{F(x)}{\dot{M}}$, the differential luminosity per unit
accretion rate $B(x)$ and  the radial profile of the temperature
of the accretion disk $T(x)$ at $\tilde{\eta}=$ 0,0.3, and
$\frac{16}{27}$, respectively.}

\end{center}
    \end{table}
A comparative study our outcome with the outcome acquired in
\cite{LUIS} shows that the percentage change of the peak values of
physical quantities $\frac{F(x)}{\dot{M}}$, $B(x)$ and $T(x)$ is
somewhat smaller than that of the values acquired in \cite{LUIS}.
We have additionally seen a conjoint impact of it in table-II
where we find that the mathematical worth of $\epsilon$ is
additionally observably  substantially  lower in this present
circumstance.

\section{Summary and conclusion}
In this article, we have studied quantum gravity corrections to
the thermal properties of a relativistic thin accretion disk
around a GUP improve Schwarzschild black hole within the IR-limit
of the asymptotic safety situation for quantum gravity. We have
calculated, precisely, the corrections to the time-averaged energy
flux, the differential luminosity at infinity, the disk
temperature, and conjointly the conversion potency of accreting
mass into radiation in comparison to the predictions of classical
general relativity theory.

We have found that an increase in the parameter $ \eta$ that
encodes the quantum effects within the GUP framework, not solely
results in a shifting of the radius of the inner fringe of the
disk and also the ISCO, toward smaller values, but, as a
consequence, we have found a tendency to rising the energy
radiated far off from the disk together with an increase in
temperature of the disk. We have noticed a bent to conjointly rise
in the differential luminosity reaching the observer far away at
infinity along with a higher conversion potency of accreting mass
into radiation. Besides, a shifting of the height of the radial
profiles the thermal properties toward smaller values of the
radial coordinate have conjointly been observed.

In \cite{LUIS}, it has been shown that experimental knowledge
reveals that this model is associated with nursing correct one at
low luminosities however it is not so well appropriate at high
luminosities, a regime that a skinny accretion disk provides a
much better description \cite{JFS, OST}. Throughout this text it
has been reinstated everywhere once more. Once again, our
investigation shows that quantum gravity within the physics of
black hole manifests itself at distances even larger than the
radius of the ISCO and do not seem to be restricted inside of the
horizon so greatly to its immediate neighborhood. Our result
encourages the study of quantum gravity effects on several
realistic black holes environments, like accretion onto quantum
improved Kerr, Kerr-Sen black hole as a varied or complementary
path to confront the predictions of asymptotic safety with
astronomical observation.

{\bf Data availability Statement:}It is a theoretical paper.
  Data used here are generated from numerical computation

\end{document}